\documentclass[conference]{IEEEtran}
\IEEEoverridecommandlockouts
\usepackage{enumitem}
\usepackage{multirow}
\usepackage{booktabs}
\usepackage{graphicx}
\usepackage{algorithm}
\usepackage{algpseudocode}
\usepackage{amsmath}
\usepackage{float}
\usepackage{subcaption}
\usepackage{amssymb}
\usepackage{array}
\usepackage{amsmath}
\usepackage{mathtools}
\usepackage[font=small]{caption}
\usepackage{cite}
\usepackage{cleveref}
\usepackage{bm}
\usepackage{xcolor}

\ifCLASSINFOpdf
\else
\fi

\begin{document}
%

\title{Fair Dynamic Spectrum Access via Fully Decentralized Multi-Agent Reinforcement Learning}



\author{Yubo Zhang*, Pedro Botelho*, Trevor Gordon*, Gil Zussman, and Igor Kadota
\IEEEcompsocitemizethanks{\IEEEcompsocthanksitem Yubo Zhang, Pedro Botelho, and Igor Kadota are with the Department of Electrical and Computer Engineering, Northwestern University, USA. Pedro Botelho is also with the Center of Excellence in Artificial Intelligence (CEIA), Brazil. Trevor Gordon and Gil Zussman are with the Department of Electrical Engineering, Columbia University, USA. This work was supported in part by the NSF under grants SES-2332054, AST-2232455, AST-2132700, and CNS-2148128 and in part by the Federal Agency and Industry Partners as Specified in the Resilient and Intelligent NextG Systems (RINGS) Program. P.\ Botelho's research was supported by CEIA and has been funded by the Acknowledge Center of Immersive Technologies (AKCIT). E-mail: yubozhang2023@u.northwestern.edu, plrbotelho.ai@gmail.com, tjg2148@columbia.edu, gil.zussman@columbia.edu, kadota@northwestern.edu. 
*These authors contributed equally to this work.}
}


%


\maketitle

\begin{abstract}
We consider a decentralized wireless network with several source-destination pairs sharing a limited number of orthogonal frequency bands. Sources learn to adapt their transmissions (specifically, their band selection strategy) over time, in a decentralized manner, without sharing information with each other. Sources can only observe the outcome of their own transmissions (i.e., success or collision), having no prior knowledge of the network size or of the transmission strategy of other sources. The goal of each source is to maximize their own throughput while striving for network-wide fairness. We propose a novel fully decentralized Reinforcement Learning (RL)-based solution that achieves fairness without coordination. The proposed Fair Share RL (FSRL) solution combines: (i)~state augmentation with a semi-adaptive time reference; (ii)~an architecture that leverages risk control and time difference likelihood; and (iii)~a fairness-driven reward structure. We evaluate FSRL in more than $50$ network settings with different number of agents,
different amounts of available spectrum, in the presence of jammers, and in an ad-hoc setting. Simulation results suggest that, when we compare FSRL with a common baseline RL algorithm from the literature, FSRL can be up to $89.0\%$ fairer (as measured by Jain's fairness index) in stringent settings with several sources and a single frequency band, and $48.1\%$ fairer on average.
\end{abstract}

\IEEEpeerreviewmaketitle

\section{Introduction}\label{sec:intro}


Future wireless applications and devices will increasingly rely on \emph{Dynamic Spectrum Access} (DSA) algorithms to effectively manage limited spectrum resources. The significance of DSA for next-generation networks has been highlighted in the National Spectrum Strategy~\cite{nationalSpectrum}. Extensive research has been conducted on developing DSA algorithms that can efficiently allocate frequency spectrum to wireless devices while minimizing harmful interference (see surveys~\cite{dsasurvey, survey3}). In recent years, \textit{Reinforcement Learning} (RL) emerged as a promising approach to enabling spectrum sharing in decentralized communication networks (see recent survey~\cite{survey2}) with sources/agents learning to make decisions over time by interacting with the environment and with other sources/agents. 

\vspace{0.05cm}
\noindent\textbf{Related Work.} \emph{Achieving fairness is a major challenge in RL-based DSA~\cite{central1,centrl2withfed,Oshri,Liew,knowactiveusers,porml,drlincn,fairdistributed}.} 
Two common approaches to achieve fair allocation of resources are: 
(i)~\textbf{centralized training}~\cite{central1,centrl2withfed,Oshri,Liew} in which all RL agents train together using a reward structure that captures network-wide fairness, thus allowing them to learn to coordinate transmissions; or  
(ii)~\textbf{information sharing}~\cite{knowactiveusers,porml,drlincn,fairdistributed} in which RL agents are allowed to share information \emph{explicitly}~\cite{knowactiveusers} or \emph{implicitly}~\cite{porml,drlincn,fairdistributed}. 
For example, the DARPA Spectrum Collaboration Challenge allowed sources/agents to explicitly share information about their future planned transmissions. 
Another example of explicit sharing is~\cite{knowactiveusers} that considers a network in which, at the end of every time slot~$t$, the centralized Access Point shares information about the outcomes of transmissions in all bands. 
An example of implicit sharing is~\cite{porml} in which agents that can sense transmissions in every frequency band and identify their source. 

Most relevant to this paper are~\cite{Oshri,Liew} which consider networks in which sources/agents can only observe the outcome of their own transmissions.  In~\cite{Oshri}, the authors consider RL agents that first train offline in a centralized manner and then train online in a decentralized manner. 
During offline training, agents learn how to coordinate transmissions. 
During online training, agents fine-tune their individual deep Q-networks (DQN). 
In~\cite{Liew}, the authors consider two distinct goals: maximizing throughput and achieving fairness. 
For maximizing throughput, the authors consider RL agents that train in a fully decentralized manner without sharing information. 
For achieving fairness, the authors consider RL agents that train in a centralized manner. 
Clearly, for both~\cite{Oshri,Liew}, centralized training is essential for achieving fairness. 
\vspace{0.05cm}
\noindent {\textbf{Main Contributions.}} In this paper, we develop a \emph{fairness-driven DSA algorithm} for decentralized communication networks in which RL agents -- called \textit{Fair Share Reinforcement Learning} (FSRL) agents -- \emph{learn/train in a decentralized manner without sharing information with each other, explicitly or implicitly}. Specifically, FSRL agents can only observe the outcomes of their own transmissions (i.e., success or collision) and they have no knowledge about the network size nor about the prior/current/future actions taken by other FSRL agents. 
To achieve fairness in a network setting with limited knowledge, we propose FSRL agents that incorporate: (i)~state augmentation with a semi-adaptive binary time reference; (ii)~an RL architecture that leverages risk control~\cite{IQN} and time difference likelihood~\cite{likelihood}; and (iii)~a novel reward structure tailored for achieving fairness without coordination. We evaluate FSRL in several network settings with different number of agents, different amounts of available spectrum, in the presence of jammers, and in an ad-hoc setting. Simulation results suggest that, when we compare FSRL with a baseline RL-based DSA algorithm from the literature~\cite{Oshri, Liew}, FSRL can be up to $89.0\%$ fairer in settings with extremely scarce resources, and $48.1\%$ fairer on average, as measured by the Jain's fairness index~\cite{jain}.


The remainder of this paper is organized as follows. 
In Sec.~\ref{sec:model}, we describe the communication network model. 
In Sec.~\ref{sec:solution}, we propose FSRL agents, describing their state, architecture, and reward. 
In Sec.~\ref{sec:experiment}, we present extensive simulation results. 
Section~\ref{sec:conclusion} concludes this paper.

\section{Decentralized Communication Network}\label{sec:model}
\begin{figure}[t]
    \centering
    \includegraphics[width=\columnwidth]{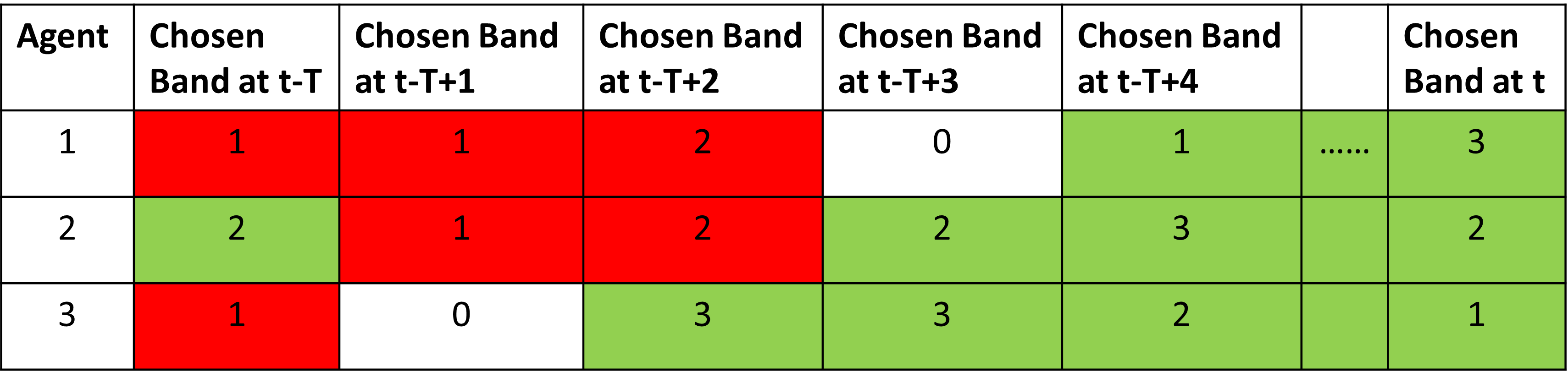}
    \caption{DSA network with $3$ source-destination pairs denoted $\{1,2,3\}$ sharing $3$ bands denoted $\{1,2,3\}$. In each slot $t$, each source $m$ transmits in band $n$ or idles (i.e., ``transmits'' in band $0$). Successful transmissions are green. Collisions are red. Idle agents are white.}
    \label{fig:model}
    \vspace{-0.4cm}
\end{figure}

We consider a wireless network composed of $M$ source-destination pairs sharing $N$ orthogonal frequency bands. 
We consider a broadcast channel\footnote{A more complex ad-hoc channel model will be discussed in Sec.~\ref{sec:ad-hoc}.} in which all sources can interfere with each other. 
We assume that sources always have packets to transmit and destinations are continuously listening to all $N$ bands. 
Let $a_m(t)\in\{0,1,\ldots,N\}$ represent the action taken by source $m\in\{1,\ldots,M\}$ in time slot $t\in\{1,\ldots,H\}$, where $H$ is the time-horizon. 
Action $a_m(t)=0$ indicates that the source idles. 
Action $a_m(t)=n$ indicates that the source transmits a packet using band $n\in\{1,\ldots,N\}$. 
Let $o_m(t)\in\{-1,0,1\}$ represent the outcome of the action taken by source $m$ in time slot $t$. 
The outcome $o_m(t)$ is revealed to each source at the end of slot $t$. 
If during slot $t$ the source idles, then $o_m(t)=0$.   
If during slot $t$ only source $m$ transmits in band $n$, then its transmission is successful ($o_m(t)=1$) and the associated destination sends a short acknowledgment to the source using the same band. 
Otherwise, if two or more sources transmit in the same band, then there is a packet collision ($o_m(t)=-1$), the associated destinations cannot decode their message, and no acknowledgment is sent. 
The transmission outcome $o_m(t)$ depends on the decisions $a_m(t)$ taken by all sources, as illustrated in Fig.~\ref{fig:model}. 
\textbf{We assume that sources cannot share information to coordinate transmissions. 
Specifically, in time slot $t$, source $m$ only knows historical information about its own decisions $\{a_m(k)\}_{k\leq t}$ and outcomes $\{o_m(k)\}_{k<t}$.
Sources have no prior knowledge about the network size $M$ nor about the prior/current/future actions taken by other sources.}

\section{Fair Share Reinforcement Learning (FSRL)}\label{sec:solution}

In this section, we describe our proposed solution to the problem of multiple sources dynamically and independently selecting actions aiming to maximize their own throughput (i.e., rate of successful transmissions) while striving for network-wide fairness. We use decentralized RL with each source $m$ running a separate FSRL agent responsible for selecting actions $a_m(t)$ over time. Next we describe the agent's state, architecture, and reward.






\subsection{Augmented State of the FSRL Agent}\label{sec:observation}

In time slot $t$, the FSRL agent associated with source~$m$ selects an action $a_m(t)$ utilizing historical information, namely actions $\{a_m(t-T),\ldots,a_m(t-1)\}$ and outcomes $\{o_m(t-T),\ldots,o_m(t-1)\}$ from the previous $T$ time slots, where $T$ is the temporal length. An example of historical information for a particular agent, for $T=5$, and using one-hot encoding to represent $a_m(t)$ is shown in the three bottom rows of Table~\ref{table:actions_success}. 

\vspace{0.05cm}
\noindent\textbf{Semi-adaptive Binary Time Reference.} We augment the FSRL agent's state with a time reference counter which represents time slot $t$ modulo $16$, i.e., $mod(t,16)$, allowing the time reference to be represented using $4$ bits. 
For example, $t=27$ gives $mod(t-1,16)=10$ which is represented by $(1010)_2$ in the top four rows in the last column in Table~\ref{table:actions_success}. Then, the augmented state at time $t$, i.e., $\bm{s}_m(t)$, is composed of actions, outcomes, and binary time references from the previous $T$ time slots. Table~\ref{table:actions_success} illustrates the augmented state of a FSRL agent at time $t=27$. The augmented state is all that a FSRL agent can observe before selecting an action.
 
\textbf{By providing a time reference to FSRL agents, we aim to facilitate their pursuit of transmission patterns.} 
For example, consider a scenario with two agents fairly sharing a single band. Each agent should follow a pattern similar to: transmit, idle, transmit, idle, and so on. With the binary time reference, agent 1 could learn to ignore the three Most Significant Bits (MSB) of the time reference, and transmit when the Least Significant Bit (LSB) is 1 and idle when the LSB is 0. 
\textbf{The binary representation of the time reference allows FSRL agents to ignore bits adaptively, which is useful for a dynamic environment where the number of agents in the network can change.}
The choice of $mod(t,16)$ limits the length of the transmission pattern to 16. In contrast, a value larger than 16 would enlarge the state space. 
%
In Sec.~\ref{sec:comparison}, we compare the performance of our FSRL solution with and without time reference and show that the time reference significantly improves performance.


\begin{table}[t]
\vspace{0.2cm}
\centering
\caption{Augmented state $\bm{s}_m(t)$ of a FSRL agent at time $t=27$ (with binary time reference with modulo $16$) for a network with $N=2$ frequency bands. MSB/LSB stands for Most/Least Significant Bit.}
\begin{tabular}{c|ccccc} 
\hline & t-5 & t-4 & t-3 & t-2 & t-1 \\ \hline 
\textbf{Binary time ref. (MSB)} & 0 & 0 & 1 & 1 & 1 \\ 
\textbf{Binary time reference } & 1 & 1 & 0 & 0 & 0 \\ 
\textbf{Binary time reference} &  1 & 1 & 0 & 0 & 1 \\
\textbf{Binary time ref. (LSB)} &  0 & 1 & 0 & 1 & 0 \\ \hline 
\textbf{Transmit in band 2} & 0 & 1 & 0 & 0 & 0 \\
\textbf{Transmit in band 1} & 0 & 0 & 1 & 1 & 0 \\
\textbf{Outcome} & 0 & -1 & -1 & 1 & 0 \\ \hline 
\end{tabular}
\label{table:actions_success}
\vspace{-0.4cm}
\end{table}

\subsection{FSRL Network Architecture}\label{sec:architecture}

\begin{figure}[t]
    \centering
    \includegraphics[width=\columnwidth]{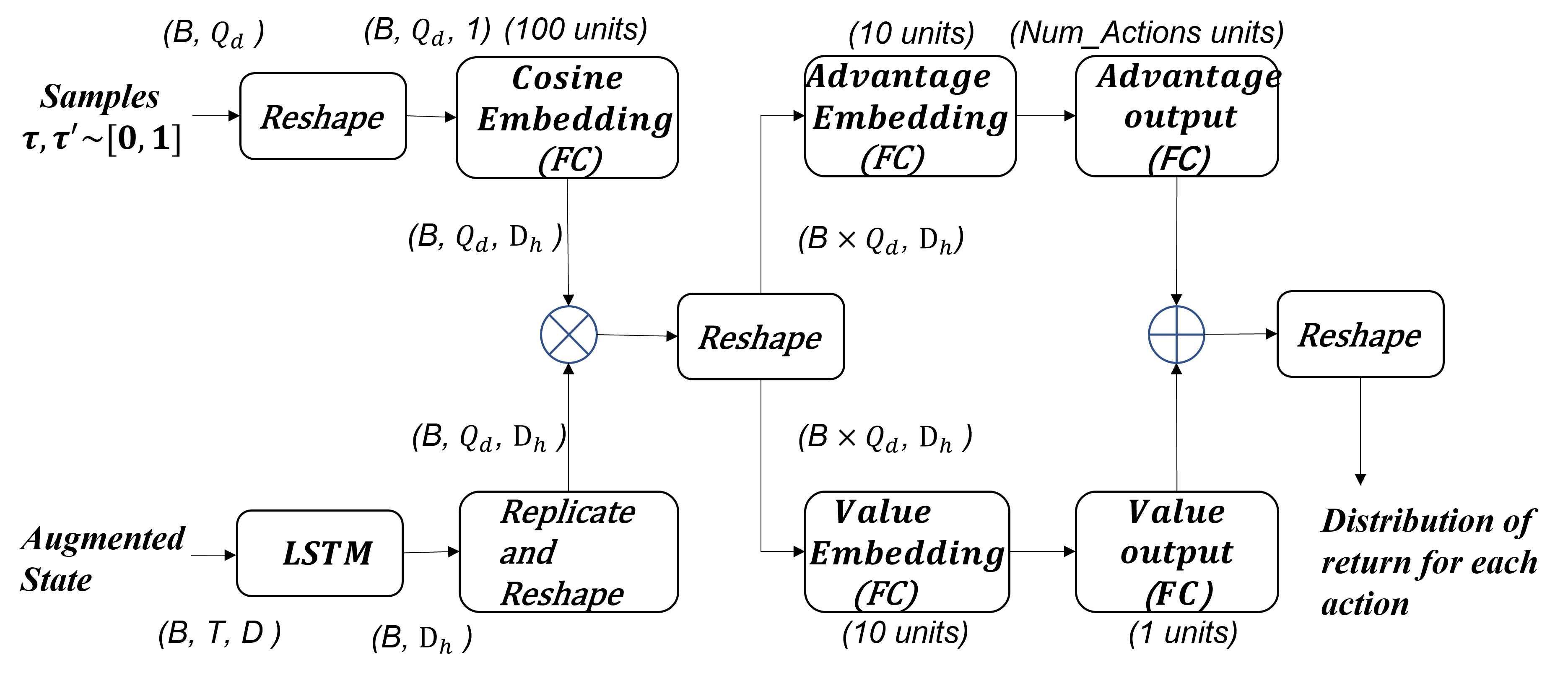}
    \caption{Architecture of each FSRL agent which integrates Dueling DQN with Distributional RL. Legend: $B$ is the batch size, $Q_d$ is the quantile dimension, $T$ is the temporal length, $D_h$ is the number of hidden units, and $D$ is the feature dimension.}
    \label{fig:architecture}
    \vspace{-0.2cm}
\end{figure}

The proposed architecture of FSRL agents is illustrated in Fig.~\ref{fig:architecture}. 
This architecture is inspired by~\cite{Oshri} which uses Dueling Deep Q Network (DDQN). 
Recall that, as described in Related Work in Sec.~\ref{sec:intro}, to achieve fairness, the DDQN solution~\cite{Oshri} relied on centralized training of all RL agents in the network. 
In this paper, \textbf{aiming to foster collaboration during a fully decentralized training process, we enhance the DDQN architecture with the Likelihood Hysteretic Implicit Quantile Network (LH-IQN) proposed in~\cite{IQN,likelihood}.} Next, we briefly introduce the DDQN architecture used in~\cite{Oshri}, then we describe our enhancements leveraging Implicit Quantile Network, Dynamic Risk, and Time Difference Likelihood. 
For reproducibility, we will share the code online prior to the conference. 

\vspace{0.05cm}
\noindent\textbf{Dueling Deep Q-Networks.} DDQNs extend the traditional Q-learning framework by decomposing the Q-value function into two components: the state value function and the advantage function. 
The advantage layer and value layer can be seen on the right side of Fig.~\ref{fig:architecture}. 
This separation allows the evaluation of the importance of states independently of the actions, which improves generalization across similar state-action pairs, especially when different actions lead to similar outcomes. 
When compared with traditional Q-learning, DDQNs have shown performance benefits~\cite{ddqn} especially in environments with large state-action spaces (similar to this paper). 

\vspace{0.05cm}
\noindent\textbf{Implicit Quantile Networks.} Distributional RL extends traditional RL by modeling the entire distribution of future returns instead of modeling only their expected value. Let the return \(Z^\pi(s, a) = r_t + \gamma r_{t+1} + \gamma^2 r_{t+2} + \dots\) represent the cumulative future reward under a given policy \(\pi\) from a given state-action pair \((s, a)\), where \(r_t\) is the immediate reward provided by the environment at time $t$ and \(\gamma\in(0,1]\) is the discount factor.  
Unlike the immediate reward \(r_t\), which is a single scalar value, the return distribution captures the variability and uncertainty of future outcomes under a given policy \(\pi\). Implicit Quantile Networks (IQN) build upon the principles of Distributional RL by estimating the return distribution through a quantile-based approach. Specifically, IQN approximates the inverse cumulative distribution function of \(Z^\pi(s, a)\), denoted as \(F_\pi^{-1}(s, a; \tau)\), for quantile fractions \(\tau \sim \mathcal{U}([0, 1])\). Hence, instead of estimating a single expected return \(Q^\pi(s, a) = \mathbb{E}[Z^\pi(s, a)]\), IQN estimates a set of quantiles that collectively represent the return distribution, thus providing a flexible and expressive framework for modeling the diverse range of possible outcomes associated with each action. 

\textbf{An important contribution of this paper is to propose an RL architecture that integrates DDQN and IQN}. 
%
Specifically, our architecture is based on a quantile-based distributional DDQN which models the return distribution \(Z^\pi(s, a)\) for each state \(s\) and action \(a\). Next, we describe the main steps of the FSRL architecture illustrated in Fig.~\ref{fig:architecture}.

The input to the FSRL architecture are: (i) the augmented state \(\bm{s}_m(t) \in \mathbb{Z}^{T \times D}\), where $D$ is the feature dimension; and (ii) the set of sampled quantile fractions $\bm{\tau}\in\mathbb{R}^{B \times Q_d \times 1}$, where each individual $\tau \sim \mathcal{U}([0, 1])\), \(B\) is the batch size, and \(Q_d\) represents the quantile dimension.
%
%
The augmented state \(\bm{s}_m(t)\) is processed by an LSTM layer to encode temporal dependencies, giving us 
$\bm{h}_t = \text{LSTM}(\bm{s}_m(t)) \in \mathbb{R}^{B \times D_h}$, 
where $D_h$ is the number of hidden units. 
The output $\bm{h}_t$ is then replicated across the quantile dimension and reshaped which yields 
$\bm{h}_t' \in \mathbb{R}^{B \times Q_d \times D_h}$.

Simultaneously, the set of sampled quantile fractions $\bm{\tau}$ are adjusted using a risk-sensitive transformation, such as:
\begin{equation}\label{eq:risk}
\bm{\tau}_{\text{distorted}} = W(\bm{\tau}) = \Phi\left(\Phi^{-1}(\bm{\tau}) + \alpha\right),
\end{equation}
where $W(\bm{\tau})$ is the Wang transformation\cite{Wang2000ACO}, \(\Phi\) is the standard normal cumulative distribution function (CDF), and $\alpha$ is a risk parameter. A positive $\alpha$ corresponds to risk-seeking behavior, while a negative $\alpha$ corresponds to risk-averse behavior. This distortion modifies the network's focus on specific parts of the distribution, such as low or high returns. 
Next, the distorted quantiles are transformed using cosine embeddings
\begin{equation}
\bm{\phi}(\bm{\tau_{\text{distorted}}}) = \cos\left(\pi \bm{\tau_{\text{distorted}}} \bm{\omega}\right) \; ,
\end{equation}
with $\bm{\omega} \in \mathbb{R}^{D_h}$ and \(\bm{\phi}(\bm{\tau_{\text{distorted}}}) \in \mathbb{R}^{B \times Q_d \times D_h}\). This embedding introduces periodicity, enhancing the representation of the quantiles. 



The distorted quantile embeddings \(\bm{\phi}(\bm{\tau}_{\text{distorted}})\) are then element-wise multiplied with the reshaped LSTM output, producing a joint representation
\begin{equation}
\bm{z}_t = \bm{\phi}(\bm{\tau}_{\text{distorted}}) \odot \bm{h}_t' \quad \mbox{where} \quad \bm{z}_t\in \mathbb{R}^{B \times Q_d \times D_h} ,
\end{equation}
which combines state and quantile information. 
The joint representation \(\bm{z}_t\) passes through fully connected layers to compute the value \(V(s)\) and advantage \(A(s,a)\) as follows
\begin{align}
V(s) &= f_v(\bm{z}_t) \in \mathbb{R}^{B \times Q_d \times 1} \\
A(s,a) &= f_a(\bm{z}_t) \in \mathbb{R}^{B \times Q_d \times |\mathcal{A}|}
\end{align}
where \(f_v(.)\) and \(f_a(.)\) are fully connected networks, and \(|\mathcal{A}|\) is the number of actions.
The distribution of returns, denoted as \(Z(s,a;\tau)\in \mathbb{R}^{B \times Q_d \times |\mathcal{A}|}\), is obtained by combining value and advantage components as follows
\begin{equation}\label{eq:z}
Z(s, a;\tau) = V(s) + \left(A(s, a) - \frac{1}{|\mathcal{A}|} \sum_{a'} A(s, a')\right),
\end{equation}
where the mean advantage is subtracted to stabilize learning. 

To minimize the Temporal Difference (TD) error, traditional DQN takes a state-action estimate from the target network and a state-action  $(s_t, a_t)$ from the current network and minimizes TD as follows $\delta = Q(s_t, a_t) - Q_{\text{targ}}$ with $Q_{\text{targ}}=r_t + \gamma \hat{Q}(s_{t+1}, \pi(s_{t+1}))$, where $r_t$ is the received reward and $\gamma$ is the discount factor.  
Similarly, given samples \( \tau , \tau' \sim U([0, 1]) \), the distributional version of TD error is defined as 
\begin{equation}\label{eq:TD}
\delta_{\tau, \tau'} = Z(s_t, a_t;\tau') - Z_{targ}(\tau) 
\end{equation}
with $Z_{targ}(\tau)=r_t + \gamma \hat{Z}_{\tau}(s_{t+1}, \pi(s_{t+1}))$, 
where \(Z(s_t, a_t;\tau')\) is given from~\eqref{eq:z} and \(\hat{Z}_{\tau}(s_{t+1}, \pi(s_{t+1}))\) denotes the distributional estimate of the next state under the greedy policy defined as \(\pi(s_{t+1}) = \arg \max_a Q(s_{t+1}, a)\).

Finally, given the distributional TD error, the IQN loss function is as follows
\begin{equation}\label{eq:IQNLoss}
    \mathcal{L}(s_t,a_t,r_t,s_{t+1}) = \frac{1}{Q_d} \sum_{i=1}^{Q_d} \sum_{j=1}^{Q_d} \rho_{\tau_i} (\delta^{\tau_i, \tau'_j})
\end{equation}
where $Q_d$ is the total number of samples \( \tau , \tau' \sim U([0, 1]) \) used to estimate the loss, and the quantile regression loss is given by
\begin{equation}\label{eq:QuantileRLoss}
    \rho_{\tau}(\delta) = (\tau - \mathbf{1}_{\delta \leq 0}) \frac{H_k(\delta)}{k}
\end{equation}
where $H_k$ is the Huber Loss with threshold $k$~\cite[Sec.~2.2]{IQN}. 

The update rule for the neural network weights and biases, represented as $\theta$, follows the equation $\theta \gets \theta - \mu_t \nabla_\theta \mathcal{L}$, where $\mu_t$ is the learning rate (dynamically adjusted according to~\eqref{eq:modifyingmu}, discussed later), and $\nabla_\theta \mathcal{L}$ is the gradient of the loss function $\mathcal{L}$ with respect to $\theta$. This process minimizes the loss by updating the weights in the direction of the negative gradient. 
To ensure stability during training, the target network with weights $\theta_{\text{target}}$ is periodically updated to match the weights of the primary network. This periodic update can be expressed as $\theta_{\text{target}} \gets \theta$, and is performed every $N$ steps. The target network provides fixed targets during loss computation, reducing instability caused by rapidly fluctuating predictions from the primary network.

\vspace{0.05cm}
\noindent\textbf{Dynamic Risk.} In settings with multiple FSRL agents that have just recently started training, many transmissions may result in collisions. In this case, agents may learn a distribution of rewards that is heavily weighted towards negative values, inducing a ``risk-averse behavior,'' e.g., remaining silent. By judiciously modifying the sampling distribution of $\tau$ and $\tau'$, it was shown in~\cite{IQN} that it is possible to emphasize higher rewards, inducing ``risk-seeking behavior,'' e.g., attempting transmissions. 
This modification of sampled $\tau,\tau'$ can be achieved by adjusting $\alpha$ in~\eqref{eq:risk} over time. 
In our simulations, we start with a risk value $\alpha=0.5$ and decrease $\alpha$ over time using a risk decay of $5e^{-4}$. 

\vspace{0.05cm}
\noindent\textbf{Time Difference Likelihood}. TDL adjusts the network's learning rate $\mu_t$ over time. Intuitively, it reduces the learning rate when it encounters agents that are in their exploration phase. To detect exploratory actions by other agents, TDL leverages samples from $Z(s_t, a_t;\tau')$ and $Z_{targ}$ to determine the likelihood $\mathcal{L}_S$ that samples are from the same distribution. Intuitively, a higher $\mathcal{L}_S$ indicates a good match between the predicted and target distributions, while a lower $\mathcal{L}_S$ suggests no overlap, reflecting poor model performance. The likelihood $\mathcal{L}_S$ is used to influence the learning rate, allowing the model to adjust its updates based on the similarity between distributions, according to
\begin{equation}\label{eq:modifyingmu}
\mu_t = \begin{cases} 
\max(\beta, \mathcal{L}_S) \cdot \bar{\mu}, & \text{if } \delta_{\tau_i, \tau'_j} \leq 0, \\
\bar{\mu}, & \text{otherwise}.
\
\end{cases}
\end{equation}
where $\bar{\mu}$ be the base learning rate (tuned for stationary environments) and $\beta$ is a threshold applied when $\mathcal{L}_S$ is too low to prevent the learning rate from becoming excessively small. This dynamic adjustment ensures that the learning process remains efficient and avoids stagnation during optimization. Details about the computation of $\mathcal{L}_S$ can be found in~\cite{likelihood}. 
\textbf{The combination of Dynamic Risk and Time Difference Likelihood is expected to significantly improve sharing of limited resources.}

\subsection{Fairness-driven Reward Structure of FSRL Agents}\label{sec:reward}

We propose a fairness-driven reward that does not require information sharing among agents. Let the reward accrued by FSRL agent $m$ at the end of time slot $t$ be as follows
\begin{equation}\label{eq:reward}
R_{m}(t) = 
\begin{cases}
0.096 \times (1 - w_{m}(t)) + \Psi_k(t) \text{ , if } o_m(t)=1 \\
-1.06 \times w_{m}(t) \text{ , if } o_m(t)=-1 \\
-0.06 \text{ , if } o_m(t)=0 \text{ and } \sum_{k=t-L}^t a_m(k)=0\\
0.0516 \text{ , otherwise}
\end{cases}
\end{equation}
where $\Psi_k(t)$ is the \emph{band sharing term} (described later in~\eqref{eq:bandsharing}) and 
\begin{equation}
w_{m}(t)=\textstyle\sum_{k=t-L}^{t-1} \mathbb{I}_{\{a_m(k)=a_m(t)\}} (2^{k-t}|o_m(k)|) 
\end{equation}
is the weight associated with agent $m$ during time slot $t$, $\mathbb{I}_{\{a_m(k)=a_m(t)\}}$ is the indicator function that is equal to 1 when the band selected at a previous time slot $k$ is the same as the band selected in time slot $t$ and equal to 0 otherwise, and
$L$ is the reward history length. We normalize $w_{m}(t)$ to the range $[0,1]$. 

\vspace{0.05cm}
\noindent\textbf{Reward Weights}. In time slot $t$, agent $m$ selects band $a_m(t)$. The weight $w_{m}(t)$ increases with the number of successful transmissions in the recent past, i.e., in previous time slots $k\in\{t-L,\ldots,t-1\}$, using the same band $a_m(t)$. 
The term $2^{k-t}$ emphasizes more recent events and de-emphasizes older events. 
A high $w_{m}(t)\in[0,1]$ reduces the reward $0.096 \times (1 - w_{m}(t))$ associated with a successful transmission at time $t$ and increases the penalty $-1.06 \times w_{m}(t)$ associated with a collision. Intuitively, this should discourage agents from transmitting uninterruptedly. A low $w_{m}(t)$ has the opposite effect, encouraging agents that have not transmitted much to do so. Notice that agents that idle receive a small reward $0.0516$, but agents that are \emph{always silent} receive a penalty $-0.06$. 

The coefficients in~\eqref{eq:reward} are obtained from hyper-parameter tuning, as part of reward engineering. The selection of reward coefficients plays a critical role in achieving desirable outcomes. While, in theory, the reward should be derived directly from the task, practitioners often find it necessary to create more detailed rewards that guide the agent's behavior~\cite{gupta2022unpacking}. In this paper, the \textbf{reward coefficients in~\eqref{eq:reward} were fine-tuned to balance the agent's incentives to transmit (in the different bands) and to idle, allowing other agents to transmit. Naturally, reward over-optimization and misgeneralization~\cite{miao2024inform} are key concerns. To demonstrate that the reward~\eqref{eq:reward} generalizes to diverse network settings, in Sec.~\ref{sec:experiment} we simulate FSRL agents (always with the same reward) in $>50$ networks with different number of agents, different amounts of available spectrum, in the presence of jammers, and in an ad-hoc setting.}

\begin{figure}[H]
    \centering
    \begin{subfigure}{\columnwidth}
        \centering
        \includegraphics[width=\columnwidth]{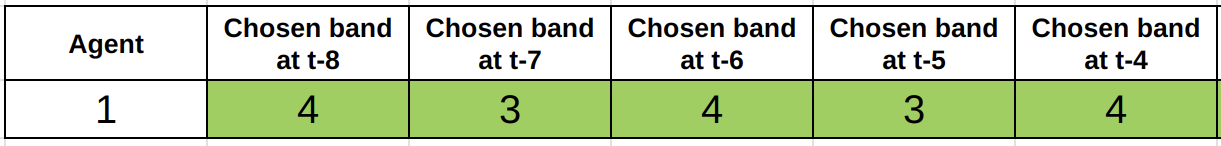}
        \vspace{-0.6cm}
        \caption{FSRL agent without band-sharing.} \label{fig:subfig1}
    \end{subfigure}
    
    
    \begin{subfigure}{\columnwidth}
        \centering
        \includegraphics[width=\columnwidth]{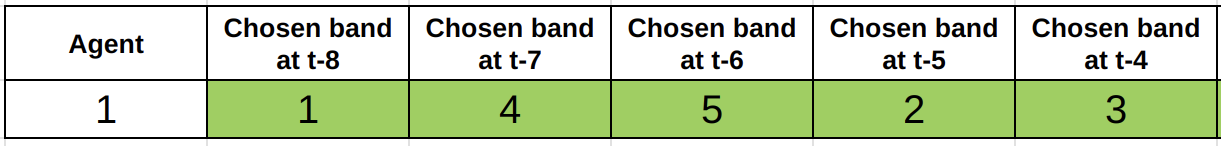}
        \vspace{-0.6cm}
        \caption{FSRL agent with band-sharing.}
        \label{fig:subfig2}
    \end{subfigure}
    \caption{Comparison of the transmissions from a FSRL agent in a network with $M=5$ agents and $N=5$ bands. (a) Shows an FSRL agent without the band-sharing term~\eqref{eq:bandsharing} in its reward~\eqref{eq:reward}. (b) Shows an FSRL agent with a reward as in~\eqref{eq:reward}.}\label{fig:bandsharing}
\end{figure}

\vspace{0.05cm}
\noindent\textbf{Band Sharing.} The band-sharing term $\Psi_m(t)$ is defined as
\begin{equation}\label{eq:bandsharing}
\Psi_m(t) = 
\begin{cases}
    \left(\frac{0.08}{1 + e^{-N+5}} + 0.12 \right)\times {G}_{m} \text{ , if } N>1 \\
    0 \text{ , otherwise}
\end{cases}
\end{equation}
where ${G}_{m}$ is a normalized geometric mean given by
\begin{equation*}
    {G}_{m} = \frac{\sqrt[N]{(B_{m,1}^L+1)(B_{m,2}^L+1)\ldots(B_{m,N}^L+1)}}{\max_{m} \left\{\sqrt[N]{(B_{m,1}^L+1)(B_{m,2}^L+1)\ldots(B_{m,N}^L+1)}\right\}}
\end{equation*}
and $B_{m,n}^L=\sum_{k=t-L}^{t-1}\mathbb{I}_{\{a_m(k)=n\}}$ represents the number of times agent $m$ transmitted in band~$n$ in the last $L$ slots. Notice that $B_{m,n}^L/L$ is the \emph{transmission rate of agent $m$ in band $n$}. Naturally, the transmission rate of agent $m$ in all bands is such that $\sum_{n=1}^N B_{m,n}^L/L \leq 1$. The normalized geometric mean\footnote{For additional information on the relationship between different types of mean, please refer to the ``mean inequality chain''.} ${G}_{m}$ tends to be \emph{larger when the values of $B_{m,n}^L$ are similar}. The normalized geometric mean uses ($B_{m,k}^L$+1) instead of $B_{m,k}^L$ to avoid persistent zeros. 
The band-sharing term $\Psi_m(t)$ increases the reward $R_{m}(t)$ in~\eqref{eq:reward} when agents spread their transmissions in different bands.
Figure~\ref{fig:bandsharing} compares the transmissions of a single FSRL agent over time slots $t$ in identical network settings in the presence/absence of the band-sharing term~\eqref{eq:bandsharing}. It is clear that the band-sharing term $\Psi_m(t)$ creates incentives for FSRL agents to spread their transmissions. An important effect of band sharing is that it makes the network more resilient to unintended interference or jamming, as highlighted in the results presented in Sec.~\ref{sec:jamming}.


\section{Experiments in Diverse Scenarios}\label{sec:experiment}
In this section, we perform an extensive evaluation of the proposed decentralized RL-based solution described in Sec.~\ref{sec:solution} in diverse scenarios. Specifically, 
in Sec.~\ref{sec:several_settings}, we show that FSRL achieves high performance in $54$ network settings with different number of agents $M$ and bands $N$; 
in Sec.~\ref{sec:comparison}, we compare the performance of FSRL with two baseline RL-based DSA algorithms; 
in Sec.~\ref{sec:jamming}, we show that FSRL agents can adapt to a jammer that enters and then leaves the network; and 
in Sec.~\ref{sec:ad-hoc}, we show that FSRL agents can adapt to ad-hoc wireless scenarios.

\begin{table}[t]
    \vspace{0.2cm}
    \centering
    \caption{Hyper-parameters used in every experiment.}
    \label{tab:hyper}
    \begin{tabular}{|lc|lc|}
        \hline
        \textbf{Parameter} & \textbf{Value} & \textbf{Parameter} & \textbf{Value} \\
        \hline
        Learning Rate ($\bar{\mu}$) & $5e^{-4}$ & Epsilon Decay & $8e^{-6}$ \\ 
        \hline
        Epsilon & $5e^{-2}$ & Risk Decay & $5e^{-4}$ \\ 
        \hline
        Risk Value ($\alpha$) & 0.5 & Temporal Length (T) & 15 \\ 
        \hline
        Buffer Size & 1500 & Update Frequency & 500 \\ 
        \hline
        Reward History Length (L) & 16 & Minimum Epsilon & $5e^{-3}$ \\ 
        \hline
        Gamma ($\gamma$) & 0.9  & Batch Size (B) & 128 \\ 
        \hline
        Quantile Dimension ($Q_d$) & 128 &  &  \\
        \hline
    \end{tabular}
    \vspace{-0.3cm}
\end{table}


\vspace{0.05cm}
\noindent\textbf{Simulation metrics}. We evaluate the DSA algorithms in terms of throughput and fairness. 
The throughput (or success rate) of agent $m$ at time $t$ is measured by 
\begin{equation}
C_{m}^{W_t}(t)=\frac{1}{W_t}\sum_{k=t-{W_t}}^{t-1}\mathbb{I}_{\{o_m(k)=1\}} \; ,     
\end{equation}
the standard deviation of agent throughput at the end of the experiment is measured by 
\begin{equation}\label{eq:throughput_std}
\sigma=\mbox{std}\{C_{m}^{W_t}(H)\} \; ,        
\end{equation}
the network throughput at the end of the experiment is measured by 
\begin{equation}\label{eq:throughput_net}
\bar{C}=\frac{1}{N}\sum_{m=1}^M C_{m}^{W_t}(H) \; .    
\end{equation}
Naturally, in the broadcast channel model, we have $\sum_{m=1}^M C_{m}^{W_t}(H)\leq N$ and $\bar{C}\in[0,1]$. The network fairness is measured using the Jain index~\cite{jain} 
\begin{equation}\label{eq:jain}
\bar{J}=\frac{\left[\sum_{m=1}^M C_{m}^{W_t}(H)\right]^2}{M\sum_{m=1}^M \left[C_{m}^{W_t}(H)\right]^2}    
\end{equation}
with a higher $\bar{J} \rightarrow 1$ indicating a fairer outcome. 

\subsection{Several Network Settings}\label{sec:several_settings}
We perform experiments for \emph{every combination} of number of source-destination pairs $M\in\{2,\ldots,10\}$ and number of bands $N\in\{1,\ldots, 10\}$ with $M \geq N$. Notice that settings with $M < N$ have spare resources and therefore are less interesting. 
\textbf{Notably, FSRL uses the same ML architecture, reward structure, and hyper-parameters described in Table~\ref{tab:hyper} in all 54 experiments.  This showcases the capability of FSRL to attain high throughput and fairness in several different scenarios without having to fine-tune the ML solution.}
The 54 experiments are conducted sequentially, without setting random seeds. Experiment results are shown ``as is,'' without replacing unfavorable results, highlighting the stability and reliability of FSRL. Repeating the same experiment multiple times and displaying averages and standard deviations is left for future work.

\begin{figure*}[t]    
    \begin{subfigure}{\textwidth}
        \centering
        \includegraphics[width=0.32\textwidth]{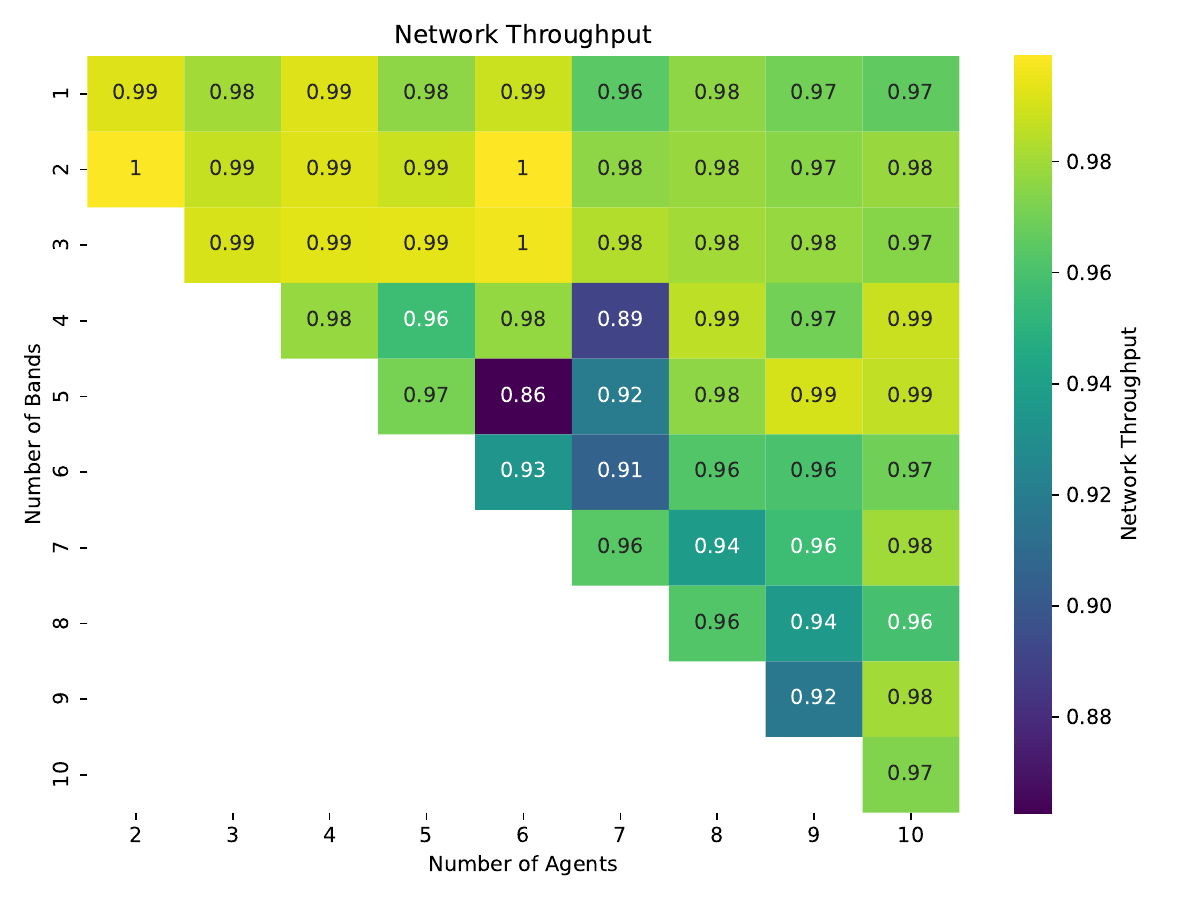}
        \includegraphics[width=0.32\textwidth]{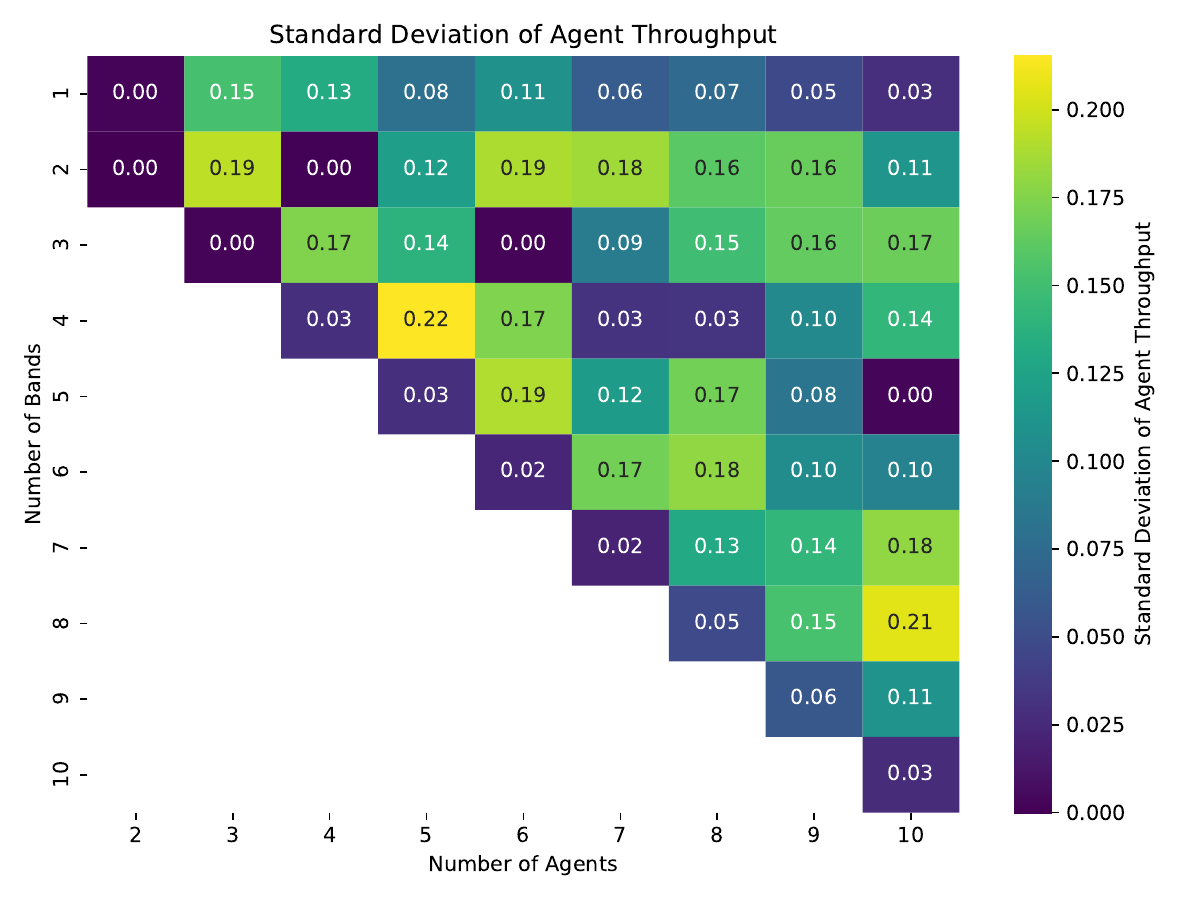}
        \includegraphics[width=0.32\textwidth]{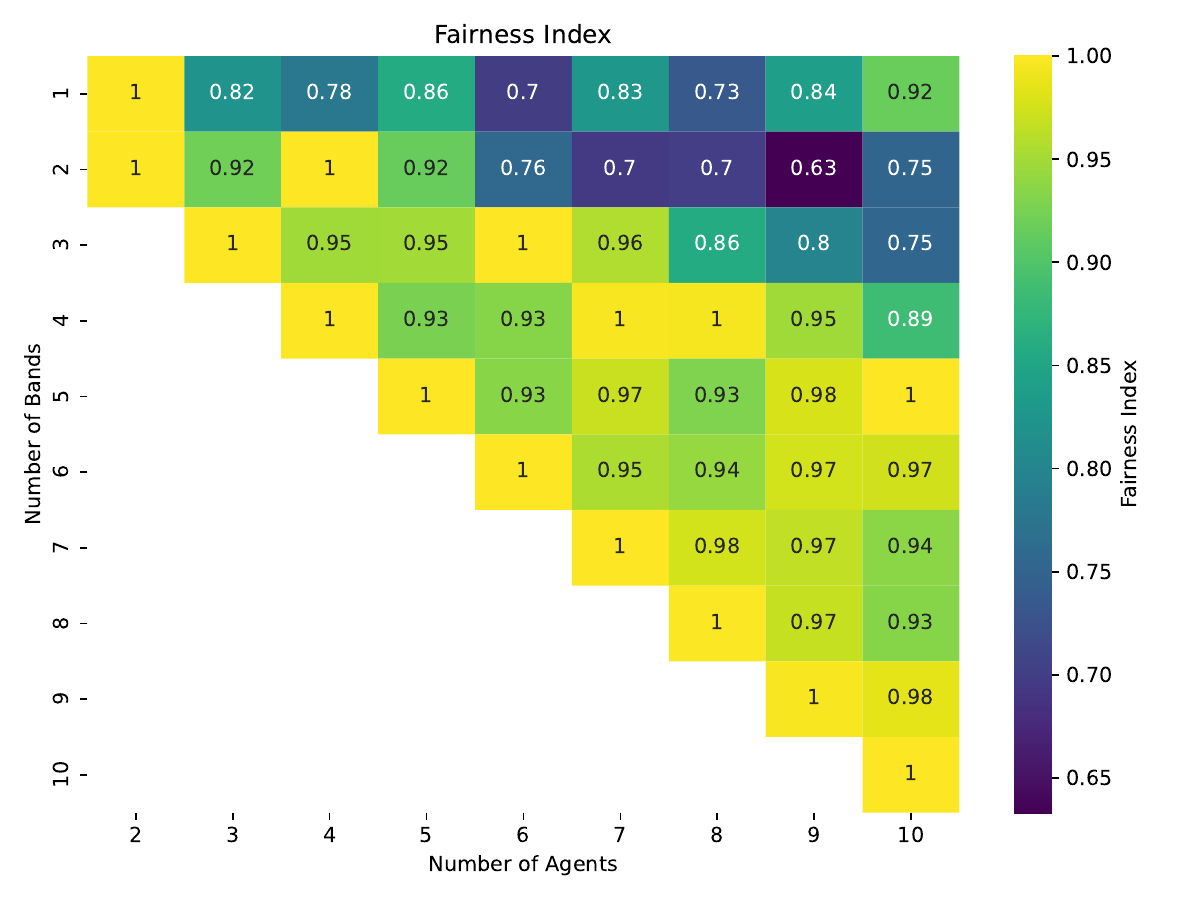}
    \end{subfigure}



    \caption{Network throughput~\eqref{eq:throughput_net}, standard deviation of agent throughput~\eqref{eq:throughput_std}, and Jain's fairness index~\eqref{eq:jain} of FSRL associated with the last $W_t=500$ time slots in diverse network settings with $M\in\{2,\ldots,10\}$ source-destination pairs and $N\in\{1,\ldots, 10\}$ frequency bands, with $M \geq N$. Notably, FSRL uses the same ML architecture, reward structure, and hyper-parameters in all $54$ experiments.}    \label{fig:overall_figure}
\end{figure*}

Figure~\ref{fig:overall_figure} displays the network fairness and throughput metrics~\eqref{eq:throughput_std}-\eqref{eq:jain} of FSRL in all $54$ experiments. It can be seen that FSRL achieves high network throughput $\bar{C}\geq 0.86$ in all settings, perfect fairness $\bar{J} = 1$ in all settings with $M=N$, almost perfect fairness $\bar{J}\geq 0.89$ for all settings with $N\geq4$, and reasonable fairness $\bar{J} \geq 0.63$ in all scenarios. The worst fairness $\bar{J} = 0.63$ occurs in the setting with $M=9$ agents and $N=2$ bands. When FSRL is compared with a baseline RL algorithm from the literature (see Table~\ref{tab:comparison}) we observe that the baseline achieves $\bar{J} = 0.22$ which is the fairness associated with 2 (out of the 9) agents uninterruptedly transmitting in the 2 available bands and the remaining 7 agents staying silent. \textbf{This comparison highlights that even the worst case scenario for FSRL still achieves reasonable fairness.}

\begin{figure}[t]
    \centering
    \begin{subfigure}{\columnwidth}
        \centering
        \includegraphics[width=\columnwidth]{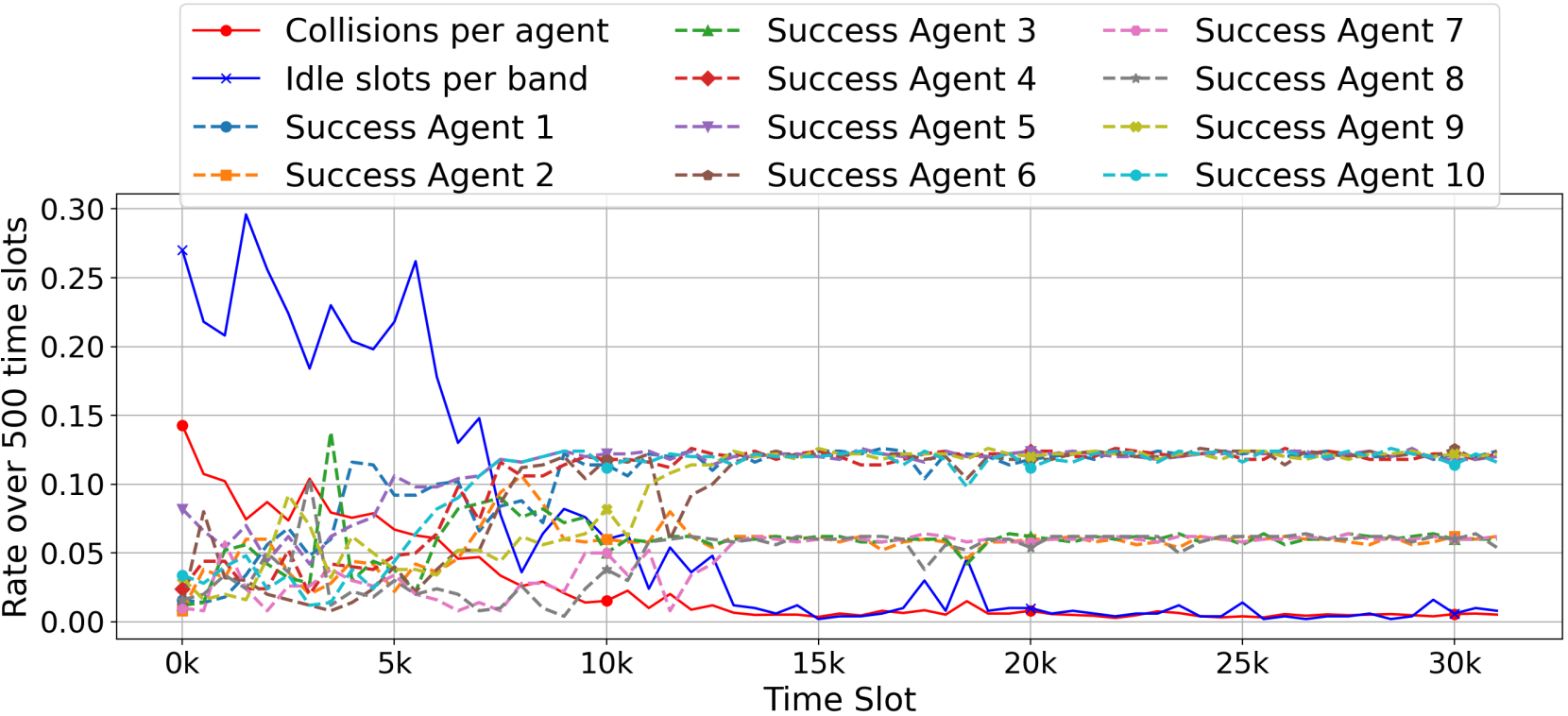}
        \vspace{-0.6cm}
        \caption{$M=10$ agents and $N=1$ band} \label{fig:subfig1}
    \end{subfigure}
    
    
    \begin{subfigure}{\columnwidth}
        \centering
        \includegraphics[width=\columnwidth]{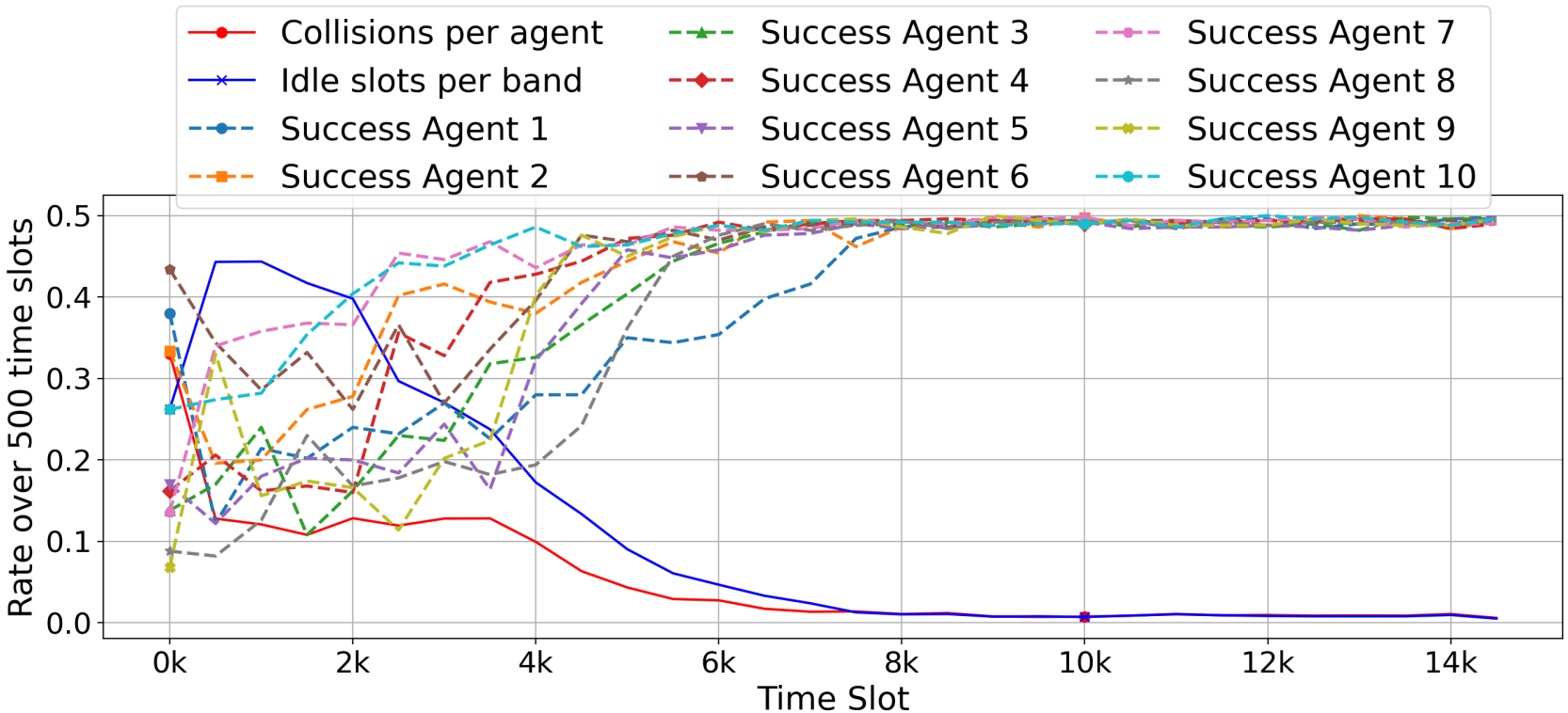}
        \vspace{-0.6cm}
        \caption{$M=10$ agents and $N=5$ bands}
        \label{fig:subfig2}
    \end{subfigure}
    
    
    \begin{subfigure}{\columnwidth}
        \centering
        \includegraphics[width=\columnwidth]{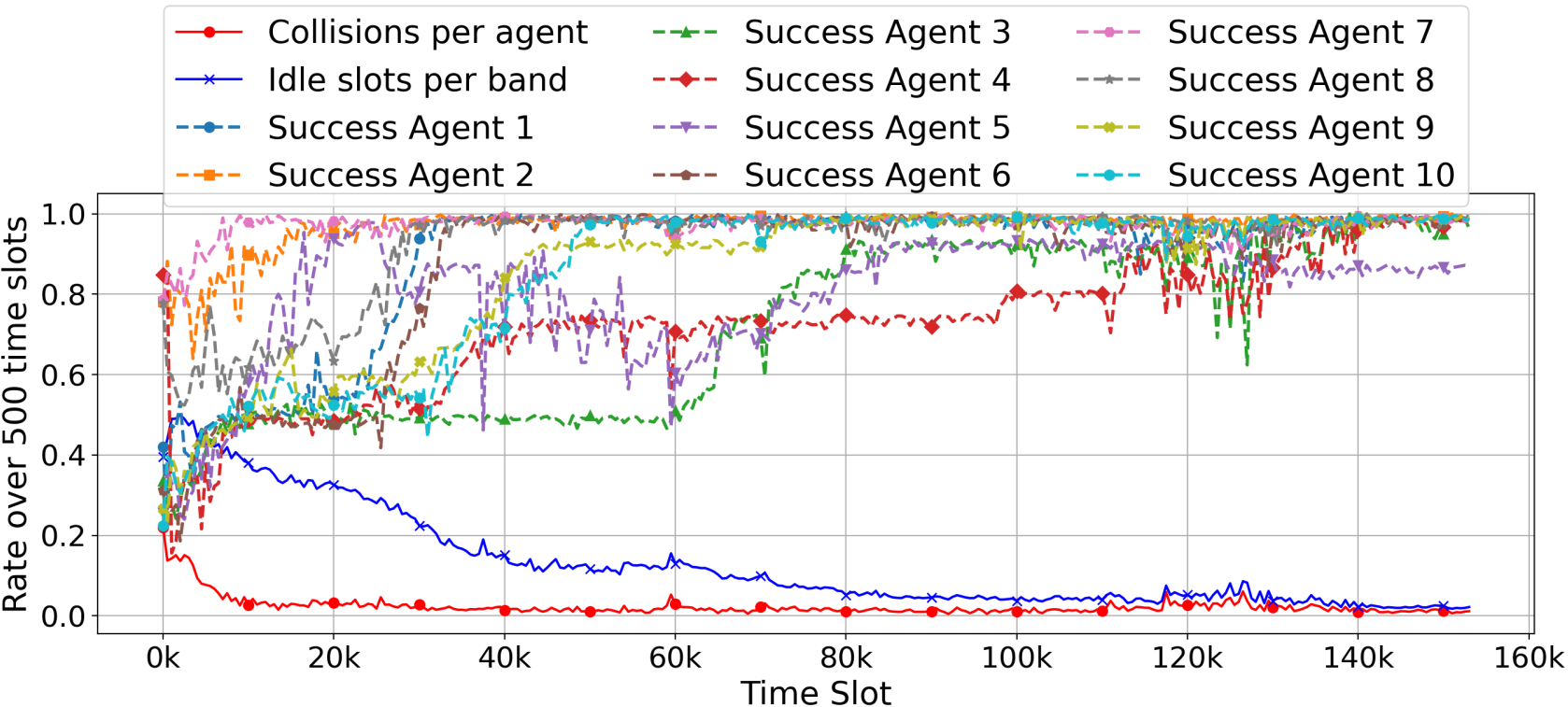}
        \vspace{-0.6cm}
        \caption{$M=10$ agents and $N=10$ bands}
        \label{fig:subfig3}
    \end{subfigure}
    \vspace{-0.5cm}
    \caption{Per agent throughput (or success rate) over time $t$ for three (out of the $54$) experiments displayed in Fig.~\ref{fig:overall_figure}.} 
    \label{fig:overall}
    \vspace{-0.2cm}
\end{figure}

Figure~\ref{fig:overall} displays the evolution of the per agent throughput (or success rate) $C_{m}^{500}(t)$ over time for three of the $54$ experiments. Notice that in all three settings the throughput of all agents converge to similar values, leading to the high fairness results shown in Figure~\ref{fig:overall_figure}. Figure~\ref{fig:overall} also displays the rate of collisions per agent and the rate of idle slots per band, both of which go to zero as time progresses, indicating that FSRL achieves high throughput. \textbf{Notably, FSRL agents achieve high throughput and fairness in a fully decentralized manner, without sharing information with each other.}

\subsection{Comparison with baseline DSA algorithms}\label{sec:comparison}

\begin{figure}[t]
    \centering
    \begin{subfigure}{\columnwidth}
        \centering
        \includegraphics[width=\columnwidth]{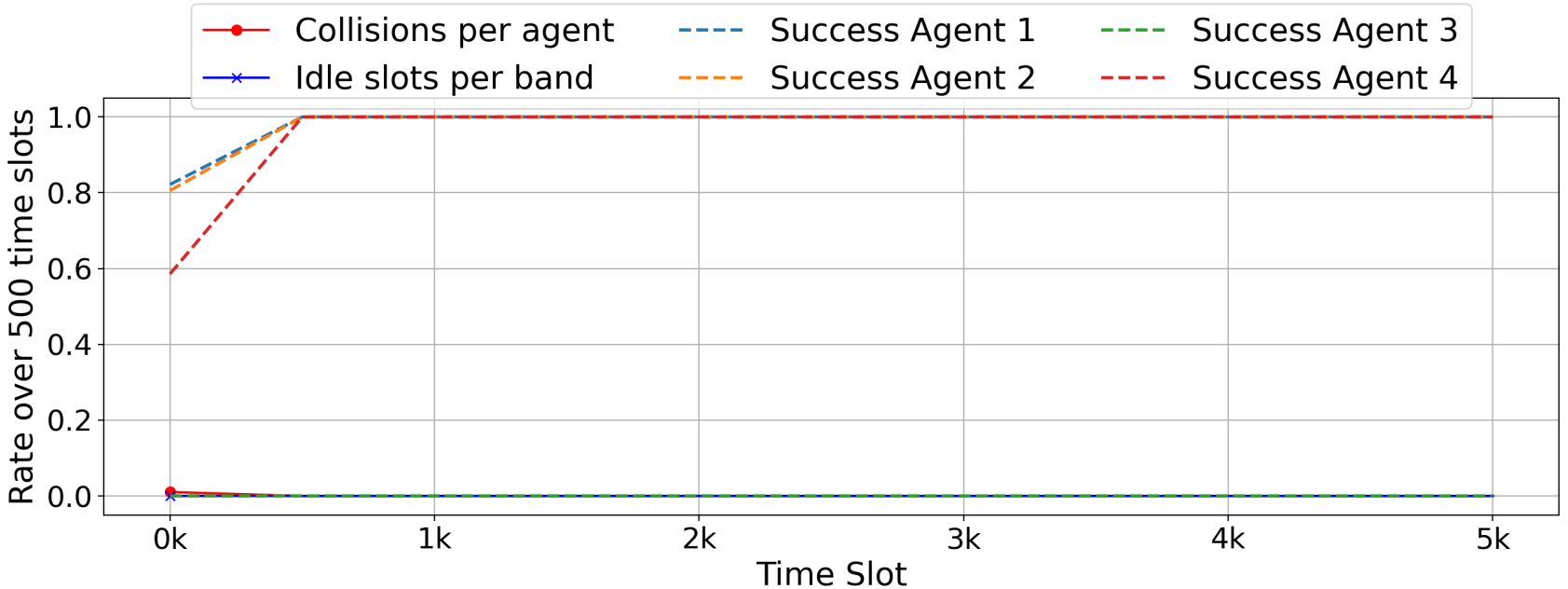}
        \vspace{-0.2cm}
        \caption{$M=4$ agents and $N=3$ bands} \label{fig:image1}
    \end{subfigure}
    
    \vspace{0.1 cm} 
    
    \begin{subfigure}{\columnwidth}
        \centering
        \includegraphics[width=\columnwidth]{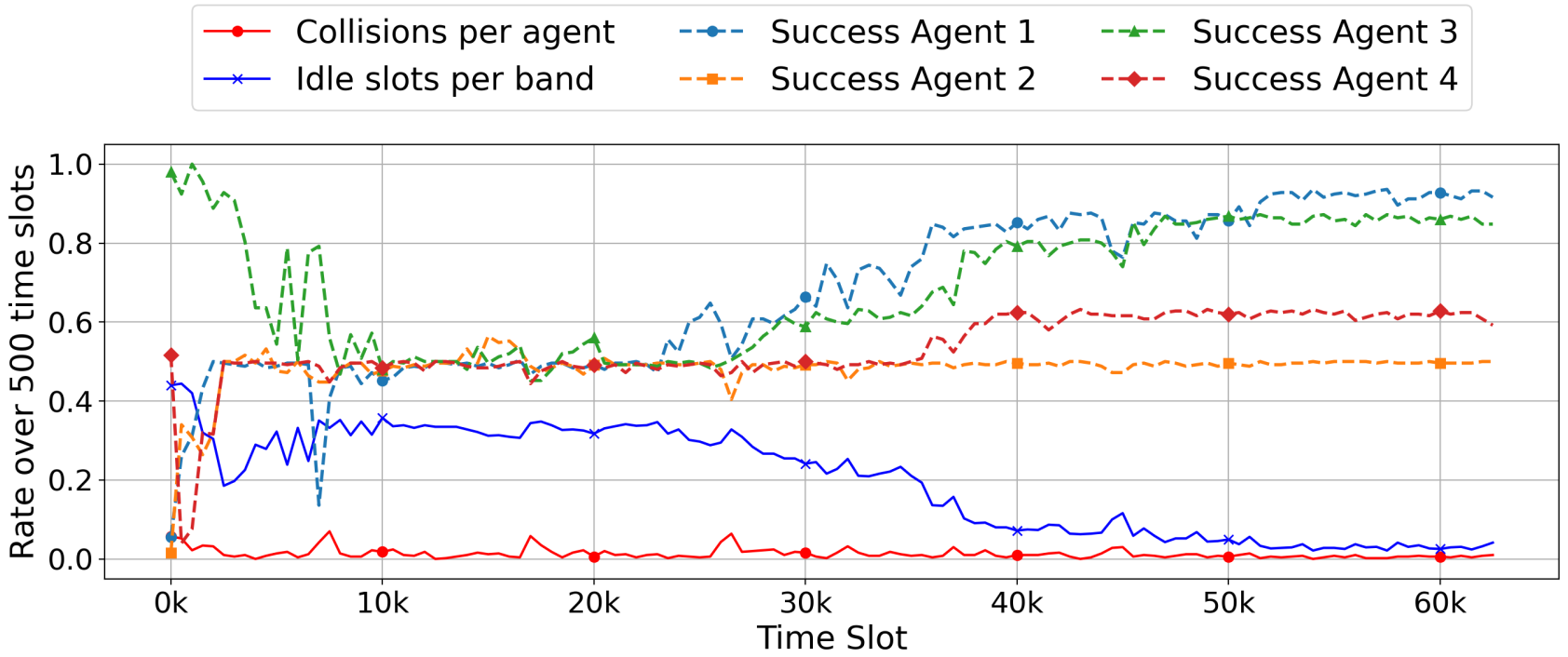}
        \vspace{-0.6cm}
        \caption{$M=4$ agents and $N=3$ bands}
        \label{fig:image2}
    \end{subfigure}
    \vspace{-0.3cm}
    \caption{Comparison of FSRL with a commonly used ML architecture and reward structure~\cite{knowactiveusers,porml,Oshri,Liew} in a network with $M=4$ agents and $N=3$ bands.} 
    \label{fig:comparison}
    \vspace{-0.4cm}
\end{figure}

An intuitive reward structure commonly used in the DSA literature~\cite{knowactiveusers,porml,Oshri,Liew} is such that RL agents accrue a fixed positive reward when their transmissions are successful and a fixed negative reward when their packets collide. 
In Figure~\ref{fig:comparison}, we compare FSRL with a solution similar to~\cite{Oshri,Liew} in which RL agents use DQN and a reward structure called Collision Penalty 1 (CP1) defined as follows
\begin{equation}\label{eq:CP1reward}
R_{m}^{CP1}(t) = 
\begin{cases}
+3 &\text{ , if } o_m(t)=1 \text{ [succ. transm.]} \\
-1 &\text{ , if } o_m(t)=-1 \text{ [collision]}\\
0 &\text{ , otherwise \text{ [idle]}}
\end{cases}
\end{equation}
Figure~\ref{fig:comparison} shows that DQN with CP1 quickly converges to an unfair outcome in which one agent remains silent, i.e., starves, throughout the experiment, while FSRL converges (after some time) to a fairer outcome in which all agents learned to share the resources. Table~\ref{tab:comparison} compares the performance of DQN with CP1, FSRL, and FSRL without binary time reference in twelve network settings. The network throughput of FSRL is on average $35.3\%$ better than FSRL without time reference, highlighting the importance of the time reference to the augmented state described in Sec.~\ref{sec:observation}. The network throughput of FSRL is on average $3.65\%$ worse than DQN with CP1. \textbf{The fairness of FSRL is on average $48.1\%$ better than DQN with CP1, highlighting the benefits of the fairness-driven reward structure} discussed in Sec.~\ref{sec:reward}.

    
    

\begin{table}[t]
\centering
\caption{Comparison of the network fairness $\bar{J}$ and throughput $\bar{C}$ for three DSA algorithms}\label{tab:comparison}
\begin{tabular}{cc|cc|cc|cc}
\hline
\multicolumn{2}{c}{Setting} & \multicolumn{2}{|c|}{DQN with CP1} & \multicolumn{2}{|c|}{FSRL w/o time ref} & \multicolumn{2}{|c}{FSRL}\\
\hline
$M$ & $N$ & $\bar{J}$ & $\bar{C}$ & $\bar{J}$ & $\bar{C}$ & $\bar{J}$ & $\bar{C}$ \\
\hline
10 & 9 & 0.90 & 1.00 & 0.98 & 0.56 & 0.98 & 0.97 \\
10 & 7 & 0.70 & 1.00 & 0.99 & 0.69 & 0.94 & 0.97 \\
10 & 5 & 0.50 & 1.00 & 0.99 & 0.99 & 1.00 & 0.99 \\
10 & 3 & 0.30 & 1.00 & 0.58 & 0.83 & 0.75 & 0.95 \\
10 & 1 & 0.10 & 1.00 & 0.48 & 0.41 & 0.91 & 0.97 \\
9 & 2 & 0.22 & 1.00 & 0.51 & 0.49 &  0.63 & 0.97 \\
8 & 2 & 0.25 & 1.00 & 0.58 & 0.74 &  0.70 & 0.98 \\
7 & 2 & 0.29 & 1.00 & 0.67 & 0.64 &  0.69 & 0.97 \\
6 & 5 & 0.83 & 1.00 & 0.99 & 0.59 &  0.94 & 0.87 \\
6 & 1 & 0.17 & 1.00 & 0.79 & 0.43 &  0.71 & 0.99 \\
5 & 4 & 0.75 & 1.00 & 0.99 & 0.60 &  0.93 & 0.96 \\
2 & 2 & 0.50 & 1.00 & 0.99 & 0.52 &  1.00 & 1.00 \\
\hline

\end{tabular}
\vspace{-0.4cm}
\end{table}




\subsection{Time-Varying Conditions: Jamming Environment}\label{sec:jamming}

\begin{figure}[t]
    \centering
    \begin{subfigure}{0.48\textwidth}
        \centering
        \includegraphics[width=\columnwidth]{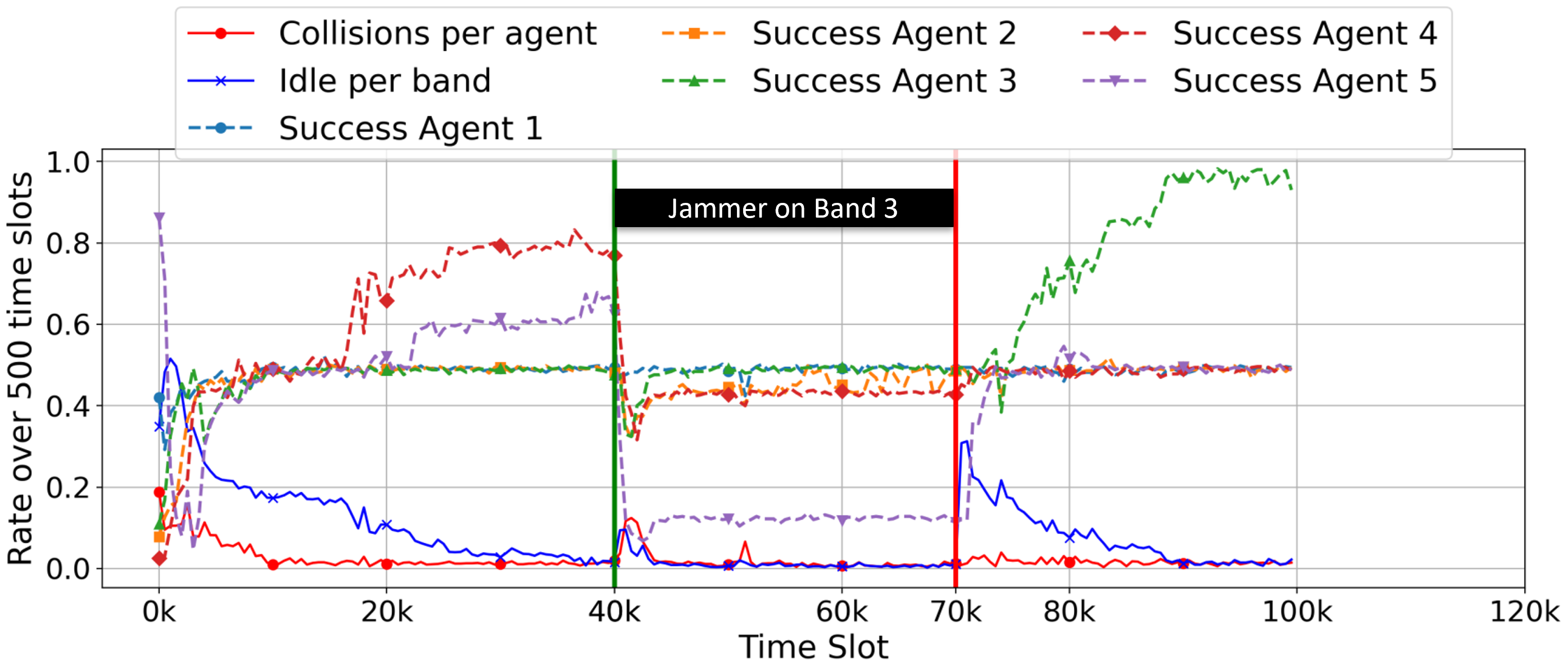}
        \vspace{-0.6cm}
        \caption{$M=5$ agents, $N=3$ bands, and jammer on band 3.}
        \label{fig:interfer53}
    \end{subfigure}
    
    
    \begin{subfigure}{0.48\textwidth}
        \centering
        \includegraphics[width=\columnwidth]{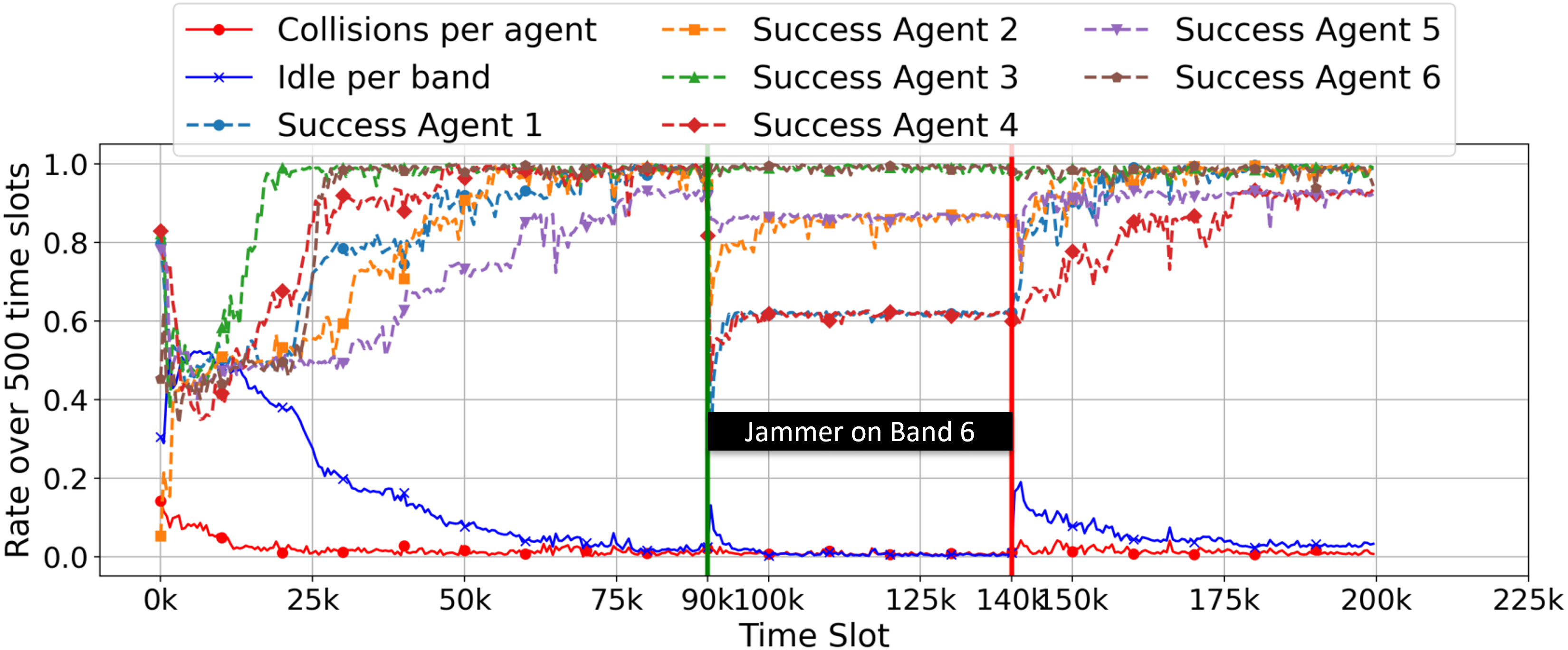}
        \vspace{-0.6cm}
        \caption{$M=6$ agents, $N=6$ bands, and jammer on band 6.}
        \label{fig:interfer66}
    \end{subfigure}
    \vspace{-0.2 cm}
    \caption{Per agent throughput over time for two experiments in which a jammer enters and then leaves the network.}
    \label{fig:interfereboth}
    \vspace{-0.2 cm}
\end{figure}

To evaluate the capability of FSRL agents to adapt to time-varying conditions, we consider a scenario in which a jammer enters and then leaves the network. Notice that adapting to a jammer that enters and remains in the network for a long time is easier than adapting to a dynamic jammer. The jammer in Fig.~\ref{fig:interfereboth}(a)/(b) enters at $t=40k$/$t=90k$ and occupies one frequency band until time slot $t=70k$/$t=140k$. As can be seen from Fig.~\ref{fig:interfereboth} and from additional experiments omitted due to space limitation, \textbf{FSRL agents maintain high throughput and fairness before, during, and after the jamming episode. The fairness-driven reward structure with a band sharing term~\eqref{eq:bandsharing} provides incentives for agents to spread their transmissions in different bands (as opposed to agents transmitting in a single band), reducing the impact of the jammer on any given agent, making it easier for the network to converge to a new fair resource allocation.} Notice that the hyper-parameters of the FSRL agents remained unchanged throughout the experiments, demonstrating the adaptability of our method to dynamic environments. In this section, FSRL uses the same ML architecture, reward structure, and hyper-parameters described in Sec.~\ref{sec:experiment} with a minimum epsilon of $0.01$.

\subsection{A More Complex Channel Model: Ad-Hoc Network}\label{sec:ad-hoc}
To evaluate the capability of FSRL agents to adapt to channels models beyond broadcast, we consider an ad-hoc network in which: 
\begin{itemize}
    \item agents only interfere with neighboring agents, i.e., agent $1$ interferes with agent $2$, agent $i$ interferes with both agents $i+1$ and $i-1$, $\forall i\in\{2,\ldots,M-1\}$, and agent $M$ interferes with agent $M-1$;
    \item a transmission from agent $i\in\{1,2,\cdots,M-1\}$ is successful only if agent $i+1$ can \emph{receive it without interference}, and a transmission from agent $M$ is successful only if agent $M-1$ can \emph{receive it without interference}.
\end{itemize}
Hence, from this ad-hoc network model, we have that a transmission from agent $i\in\{1,2\ldots,M-1\}$ in band $n$ is successful only if neither agent $i+1$ nor agent $i+2$ transmit in the same band $n$, and a transmission from agent $M$ in band $n$ is successful only if neither agent $M-1$ nor agent $M-2$ transmit in the same band $n$. Notice that, in this section, FSRL uses the same ML architecture, reward structure, and hyper-parameters described in Sec.~\ref{sec:experiment} with an initial epsilon of $0.4$, an epsilon decay of $1e^{-4}$, and a minimum epsilon of $0$. \textbf{FSRL seamlessly adapts to this new wireless channel model, suggesting that it should also be able to adapt to other more complex channel models}. 

\begin{figure}[t]
    \centering
    

    \begin{subfigure}{\columnwidth}
        \centering
        \includegraphics[width=1\columnwidth]{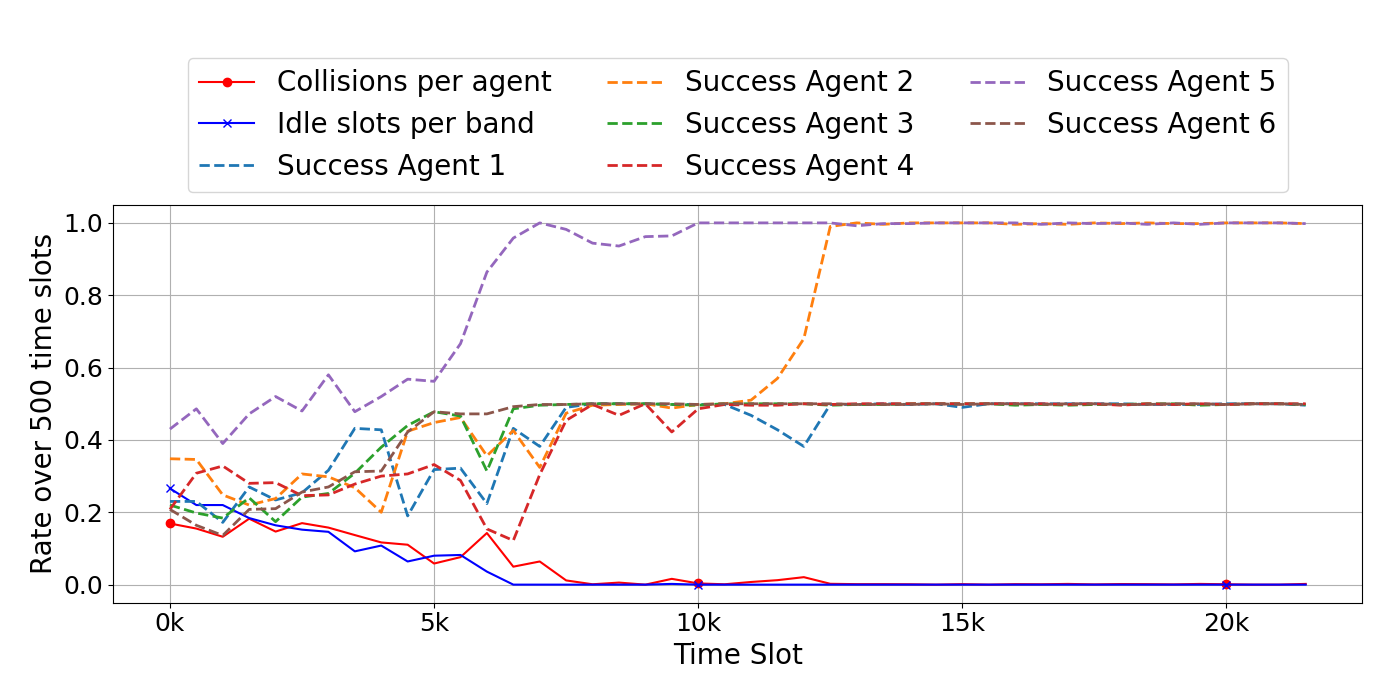}\vspace{-0.1 cm}
        \caption{$M=6$ agents and $N=2$ bands}
        \label{fig:hoc62}
    \end{subfigure}

    \vspace{-0.1 cm} 
    \begin{subfigure}{\columnwidth}
        \centering
        \includegraphics[width=0.9\columnwidth]{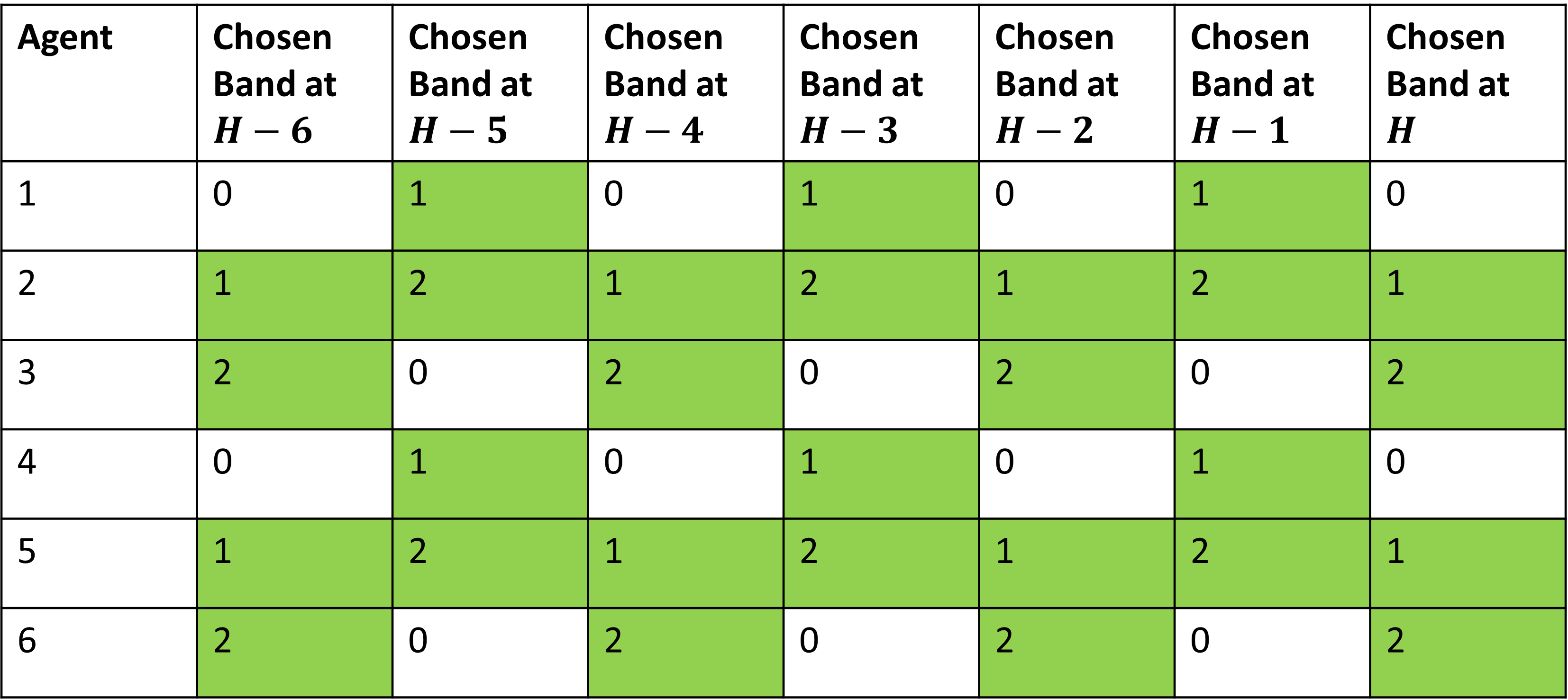}\vspace{-0.1 cm}
        \caption{$M=6$ agents and $N=2$ bands in the last $7$ time slots, where $H=22k$ is the last time slot.}
        \label{fig:hoc63}
    \end{subfigure}
    
    \caption{Per agent throughput over time for two experiments in an ad-hoc network in which agents only interfere with neighboring agents.}
    \label{fig:ad-hoc}
    \vspace{-0.2 cm}
\end{figure}

In Fig.~\ref{fig:ad-hoc} and in additional experiments with $M\in\{4,5,6\}$ and $N\in\{1,2\}$ omitted due to space limitation, we can see that FSRL agents learn to share the spectrum in this ad-hoc scenario. Summing the per agent success rate (i.e., throughput) in the last $500$ slots in Fig.~\ref{fig:hoc62}, we can see that $\sum_{m=1}^M C_{m}^{500}(H)=4>N$, indicating that FSRL agents are taking advantage of the localized interference of ad-hoc channels to transmit more often than in networks with broadcast channels. In Fig.~\ref{fig:hoc63}, we show the transmission patterns of the six agents. Notice that adding any transmissions during idle slots would result in a collision.

\section{Conclusion}\label{sec:conclusion}


In this paper, we proposed a fairness-driven DSA algorithm in which FSRL agents train in a decentralized manner without sharing information with each other. We evaluate our DSA algorithm in several network settings with different number of agents, different amounts of available frequency bands, in the presence of jammers, and in an ad-hoc setting. Simulation results suggest that, when compared with a baseline algorithm from the literature~\cite{Oshri, Liew}, FSRL can be up to $89.0\%$ fairer in settings with extremely scarce resources, and $48.1\%$ fairer on average. Furthermore, simulation results show that FSRL can achieve fairness in the presence of jammers and in ad-hoc settings. Interesting extensions include consideration of pre-training on the convergence times of FSRL and consideration of unreliable wireless channels, source mobility, and time-varying traffic loads.

\bibliographystyle{IEEEtran} 
\bibliography{reference}

\end{document}